\documentclass[final,3p,times]{elsarticle}

\usepackage[T1]{fontenc}
\usepackage{amsmath,amssymb,amsthm}
\usepackage{booktabs,longtable,array}
\usepackage[hidelinks]{hyperref}
\newtheorem{theorem}{Theorem}[section]

\newtheorem{proposition}[theorem]{Proposition}

\newtheorem{remark}[theorem]{Remark}
\newtheorem{definition}[theorem]{Definition}

\journal{Discrete Applied Mathematics}
\begin{document}
\begin{frontmatter}
\title{Two-Dimensional Search for a Moving Target on Rectangular and Cylindrical Grids}
\author{Alexey Lebedev$^1$}
\author{Vladimir Lebedev$^1$}
\author{Christian Deppe$^2$}
\address{$^1$ Higher School of Modern Mathematics, Moscow Institute of Physics and Technology, Moscow, Russia}
\address{$^2$TU Braunschweig, Institute for Information Technology, Braunschweig, Germany}

\begin{abstract}
We study how precisely a moving target can be localized on
finite rectangular grids and their cylindrical counterparts.
Each query asks whether the target lies in a chosen
subset of vertices and receives a noiseless answer. Queries may depend
on previous answers. Between successive queries, the target may stay
in place or move along one horizontal or vertical edge.
Accuracy is measured by the number of positions still possible
immediately after the final answer, with no subsequent movement.
A strategy achieves accuracy $s$ if it always terminates with at
most $s$ possible positions, whatever admissible trajectory the
target follows. Thus $s=1$ means exact localization. We denote by
$s^*(M,N)$ the smallest value of $s$ achievable by a finite
adaptive strategy on a finite grid.
Our main result shows that every finite rectangular grid admits
a strategy that localizes the target to at most eight positions,
regardless of the grid dimensions: $s^*(M,N)\leq8$. The construction
uses nine local query patterns and their vertical reflections.
Its termination follows from the progress these patterns force
toward the finite grid boundaries.
We also prove the dimension-dependent bound $s^*(M,N)\leq\min\{M+1,N+1\}$. 
For finite cylindrical grids $P_M\square C_N$, where the
columns are periodic but the first and last rows remain boundaries,
we introduce the corresponding accuracy parameter and state the
proposed dimension-independent bound
$s^*_{\mathrm{cyl}}(M,N)\leq9$. 
\end{abstract}

\end{frontmatter}
\section{Introduction}

Combinatorial search concerns the identification or localization of an
unknown object through successive queries. For a stationary object,
subset tests lead naturally to separating systems and related
information-theoretic questions. Katona's work on restricted
separating systems \cite{Katona1966}, his survey of combinatorial
search problems \cite{Katona1973}, and his discussion of R\'enyi's
contributions \cite{Katona1991} form part of the classical foundation
of this subject. General treatments of combinatorial search include
\cite{AhlswedeWegener1979,Aigner1988}.

When the object can change its position during the search, information
acquired through a test may be partly lost before the next one.
Moving-target search already appears in Koopman's probabilistic
formulation of search and screening \cite{Koopman1946}. A combinatorial
model of \emph{two-sided search} on graphs was developed by Aydinian,
Cicalese, Deppe, and V.~Lebedev
\cite{AydinianEtAl2016,AydinianEtAl2018}. In that model the searcher
queries subsets of the vertices, while the unobserved target may move
along an edge between tests. A.~Lebedev and V.~Lebedev investigated the related problem of minimizing the total cardinality of tests when searching for a moving element \cite{LebedevLebedev2019}. Faster targets and the relation between adaptive and non-adaptive tests have been investigated by A.~Lebedev and Deppe \cite{LebedevDeppe2024}. Other graph-search games also involve
hidden moving objects. For example, Beluhov and Kolev
\cite{BeluhovKolev2017} studied a chase game that includes rectangular
grids, but with a different query and capture rule. Such differences in
observations and success criteria are important when comparing results.

We investigate the finite rectangular grid
$G_{M,N}=P_M\square P_N$, with $MN$ possible target positions.
Before each binary test, the target may stay or traverse one grid edge.
The searcher may query an arbitrary subset of the vertices and adapt
subsequent queries to previous answers. We measure the remaining
uncertainty immediately after the final answer, with no subsequent
movement. With the complete grid as initial uncertainty, the same
answer histories arise if movement is placed after each nonterminal
test instead. The terminal accuracy criterion used here differs from
the one in \cite{AydinianEtAl2016,AydinianEtAl2018,LebedevDeppe2024},
which includes a target movement after the final test. In our notation,
that criterion measures $|\Gamma(D_\tau)|$, whereas we measure
$|D_\tau|$. The accuracy parameters of the two models must therefore
be distinguished. In contrast to a stationary search problem, the effectiveness of
a query depends on the geometry of the resulting candidate set, since
movement expands this set through its graph neighborhood. In two
dimensions, horizontal and vertical expansion interact, while the
finite boundary may restrict both.

An application motivating this abstraction is adaptive localization
and tracking in wireless systems employing \emph{reconfigurable
intelligent surfaces} (RISs). These surfaces can modify radio
propagation and are being investigated for wireless sensing and
localization; an overview is given by Zhang et al.
\cite{ZhangEtAl2022RIS}. RIS-assisted hierarchical beam training uses
successively refined spatial or angular probing patterns to reduce
training overhead \cite{WangEtAl2023RIS}. Moreover, RIS-aided user
localization and tracking have been studied with explicit mobility
models, for example by Teng et al. \cite{TengEtAl2022RIS} and
Keykhosravi et al. \cite{KeykhosraviEtAl2022RIS}.

The connection is particularly transparent in a discretized service
area. Each grid vertex may represent a possible user location, and
with a suitable spatial discretization and sampling interval,
mobility can be idealized by a stay-or-one-edge constraint. A sequence of suitably designed sensing
configurations can then progressively reduce the set of plausible
locations, while user mobility enlarges it again before the next
observation. The binary subset queries in the present paper idealize
this process: we assume noiseless answers and permit arbitrary test
subsets. A physical RIS or beam pattern generally cannot implement
all such subsets exactly, and practical measurements are affected by
noise, multipath, hardware restrictions, and finite angular
resolution. Thus our results provide combinatorial benchmarks for
mobility-aware adaptive localization, not performance guarantees for
a particular RIS architecture.

We study the optimal localization accuracy. For finite grids the
quantity $s^*(M,N)$ denotes the smallest worst-case number of
possible positions at termination achievable by a finite adaptive
strategy. We also consider the \emph{finite cylindrical grid}
$G^{\mathrm{cyl}}_{M,N}=P_M\square C_N$. In this model the column
coordinate is periodic: a target in column~$N$ can move directly
to column~$1$, and conversely. The number of columns remains
finite and known, and the top and bottom boundaries are unchanged.
We use the same accuracy criterion, with the candidate set counted
immediately after the final answer.

Our main contribution is the universal, dimension-independent bound
\[
 \boxed{s^*(M,N)\leq8}\qquad(M,N\geq1).
\]
We prove it directly using a nine-state adaptive strategy together with vertically reflected
configurations. An initial reduction confines the uncertainty to
four consecutive complete rows, after which a staircase query
enters the matrix strategy. The leftmost symbol of each matrix
row is repeated up to the physical left boundary. Both answers
in a distinguished state decrease the horizontal origin by one,
possibly after a vertical reflection. Every other directed cycle
has a nonzero vertical drift. The finite boundaries thus force
termination without a column-insertion induction.

An elementary adaptive diagonal sweep also yields the
dimension-dependent accuracy bound
\[
 s^*(M,N)\leq\min\{M+1,N+1\}.
\]
% Together with lower bounds and subgrid monotonicity, this
% construction determines $s^*(3,N)=4$ for $N\geq3$ and
% $s^*(4,N)=5$ for $N\geq4$, and attains accuracy seven for
% every six-column grid.

% The converse results highlight the role of geometric expansion.
% A cardinality criterion based on closed neighborhoods, together
% with finite-grid isoperimetric arguments, yields the lower bounds.
% In particular, we obtain
% \[
%  5\leq s^*(M,N)\leq\min\{8,M+1,N+1\}\qquad(M,N\geq4).
% \]
% The sharpness of the universal bound remains open for larger
% grids.

For finite cylindrical grids, Section~\ref{sec:cylindrical-grid}
introduces the graph, specifies the permissible movements, and
defines the accuracy using the same terminal timing as for
rectangular grids. In this model the column indices are cyclic,
while the upper and lower boundaries remain present. We state the
proposed universal estimate
\[
 s^*_{\mathrm{cyl}}(M,N)\leq9.
\]
Its proof is deferred. In particular, the planned analysis must
account for the periodic horizontal adjacency and for candidate
positions represented by zeros outside the printed matrix window.

Section~\ref{sec:model} specifies the model, and
Section~\ref{sec:row-reduction} develops the initial row reduction.
Section~\ref{sec:local-strategy} establishes universal accuracy eight
for finite rectangles. Section~\ref{sec:cylindrical-grid}
discusses the finite cylinder, and Section~\ref{sec:conclusion}
collects the conclusions and open questions. The complete proof of
the rectangular accuracy-eight result appears directly in
Section~\ref{sec:local-strategy}.

\section{Model and Definitions}\label{sec:model}

Let $M,N\geq1$, $[k]=\{1,\ldots,k\}$, and let
$G_{M,N}=P_M\square P_N$ be the rectangular grid with vertex set
$V_{M,N}=[M]\times[N]$. Distinct vertices $(r,c)$ and $(r',c')$
are adjacent precisely when $|r-r'|+|c-c'|=1$.
We write $R_r=\{r\}\times[N]$, $K_c=[M]\times\{c\}$,
$R_{[a,b]}=\{a,\ldots,b\}\times[N]$ and
$K_{[a,b]}=[M]\times\{a,\ldots,b\}$ for complete rows, columns,
and consecutive row or column blocks.

The target follows an admissible trajectory $v_0,v_1,\ldots$
with $\operatorname{dist}_{G_{M,N}}(v_{i-1},v_i)\leq1$.
Thus remaining in place is permitted, but diagonal motion is not.
For $A\subseteq V_{M,N}$, its closed one-step neighborhood is
\[
\Gamma(A)=\{v\in V_{M,N}:\operatorname{dist}_{G_{M,N}}(v,u)\leq1
\text{ for some }u\in A\}.
\]
Neighborhoods are always taken within the finite grid, and
$A\subseteq\Gamma(A)$.
Initially every vertex is possible, so $D_0=V_{M,N}$.

In round $i$, the target first moves; the searcher then queries
$T_i\subseteq V_{M,N}$ and receives the answer
$y_i=\mathbf{1}_{\{v_i\in T_i\}}$.
A deterministic adaptive strategy selects $T_i$ from the preceding
answers $y_1,\ldots,y_{i-1}$ and may terminate after any answer.
Tests may be arbitrary subsets, not necessarily connected or rectangular.

For a feasible answer sequence let $C_i$ and $D_i$ denote the
candidate sets immediately \emph{before} test $i$ (after movement)
and immediately \emph{after} its answer, respectively. They obey
\begin{equation}\label{eq:recursion}
C_i=\Gamma(D_{i-1}),\qquad
D_i=\begin{cases}
C_i\cap T_i,& y_i=1,\\
C_i\setminus T_i,& y_i=0.
\end{cases}
\end{equation}
In particular, $C_1=V_{M,N}$. An answer is feasible exactly when
the resulting $D_i$ is nonempty. Equivalently, a query may realize
any partition of the current candidate set.

\begin{definition}[Success and accuracy]
A strategy is called finite if its feasible decision tree is finite.
For a positive integer $s$, a finite strategy is
$(M,N,s)$-successful if, for every admissible target trajectory and
feasible answer sequence, it terminates at a stopping time $\tau$
with $|D_\tau|\leq s$. The final localization set is $D_\tau$:
no movement follows the final answer.
\end{definition}

\begin{definition}[Optimal localization accuracy]
The optimal adaptive accuracy is
\[
s^*(M,N)=\min\{s\in\mathbb N_{\geq1}:
\text{a finite adaptive }(M,N,s)\text{-successful strategy exists}\}.
\]
\end{definition}

For a finite cylindrical grid, the same candidate-set recursion
and terminal timing are used, but neighborhoods include horizontal
moves between the first and last columns; see
Definition~\ref{def:cylindrical-grid}.

\begin{remark}[Terminal timing]\label{rem:terminal-timing}
All accuracy statements in this paper concern the candidate set
$D_\tau$ immediately after the final answer. In the post-movement
model of \cite{AydinianEtAl2016,AydinianEtAl2018,LebedevDeppe2024},
the terminal uncertainty is instead $\Gamma(D_\tau)$.
These cardinalities can differ even on a terminal branch of our
construction. For example, the positive answer in matrix $A$ of Section~\ref{sec:local-strategy} retains eight vertices. In the
interior of a sufficiently large grid, their closed neighborhood
contains nineteen vertices. Thus
Theorem~\ref{thm:accuracy-eight} guarantees $|D_\tau|\leq8$;
the post-movement criterion would require a bound on
$|\Gamma(D_\tau)|$.
\end{remark}

\section{Initial Row Reduction}\label{sec:row-reduction}

We first reduce the uncertainty to two adjacent complete rows. This
initial step is used to enter the universal accuracy-eight strategy
of Section~\ref{sec:local-strategy}.

\subsection{Reduction to two adjacent rows}

The following initial phase localizes the target to two
adjacent complete rows. Its tests use complete rows, so the column
coordinate remains unrestricted. The construction is symmetric
about the middle of the grid.

\begin{proposition}[Reduction to two adjacent rows]\label{prop:two-row-reduction}
For every $M\geq2$ and $N\geq1$, starting from
$D_0=[M]\times[N]$, an adaptive strategy reaches
\[
D_\tau=\{r,r+1\}\times[N],\qquad 1\leq r<M,
\]
after finitely many tests.
\end{proposition}
\begin{proof}
Put $m=\lfloor M/2\rfloor$ and first query
$T_1=[m]\times[N]$. Because the initial candidate set is the
entire grid, the first answer leaves either
$D_1=R_{[1,m]}$ or $D_1=R_{[m+1,M]}$.

If $D_1=R_{[1,m]}$, successively query
\[
Q_j=R_{[m-j+1,m-j+2]},\qquad j=1,\ldots,m,
\]
stopping at the first positive answer. Inductively, provided the
preceding $Q$-answers were all negative, the candidate set immediately
before $Q_j$ is
\[
C_{j+1}=R_{[1,m-j+2]}.
\]
A positive answer therefore leaves exactly $D_{j+1}=Q_j$;
a negative answer, when feasible, leaves
$D_{j+1}=R_{[1,m-j]}$, establishing the induction after the next
movement. At $j=m$, the query equals the entire candidate set
$R_{[1,2]}$, so a positive answer is unavoidable.

If $D_1=R_{[m+1,M]}$, use instead
\[
Q_j=R_{[m+j-1,m+j]},\qquad j=1,\ldots,M-m.
\]
After $j-1$ negative answers, the set immediately before $Q_j$ is
\[
C_{j+1}=R_{[m+j-1,M]}.
\]
A positive answer again leaves exactly the two rows $Q_j$;
a negative answer, when feasible, leaves
$D_{j+1}=R_{[m+j+1,M]}$. The last query, at $j=M-m$, contains
the entire remaining candidate set $R_{[M-1,M]}$, and hence must
receive a positive answer.

Thus every feasible answer history reaches two adjacent complete
rows. The same argument includes $M=2$.
\end{proof}

\subsection{The next movement and boundary cases}\label{sec:row-boundary}

After this phase, $D_\tau=\{r,r+1\}\times[N]$ for some
$1\leq r<M$. Since both rows contain every column, horizontal
movement does not enlarge the candidate set. The next movement,
if the search continues, gives exactly
\begin{equation}\label{eq:two-row-neighborhood}
C_{\tau+1}=\Gamma(D_\tau)
=R_{[\max\{1,r-1\},\,\min\{M,r+2\}]}.
\end{equation}
For $2\leq r\leq M-2$ this consists of four consecutive complete
rows. At the upper or lower boundary it consists of at most three
rows (and only two when $M=2$).

We use a four-row virtual template and intersect it with the
physical grid. Its first row may lie outside $[M]$. All later tests
keep the reference coordinates prescribed by their answer history;
they are not re-anchored to the first occupied physical row.
Neighborhood monotonicity then preserves the candidate-set
containments, including the cases $M=2,3$. 
%The formal boundary truncation argument is also recorded in Section~\ref{sec:global-reduction}.

\section{Universal Accuracy Eight on Finite Grids}\label{sec:local-strategy}

This section gives the complete finite-grid construction and proof
without referring to a matrix appendix. The strategy uses nine
explicit query patterns, denoted by $A,\ldots,I$, and the patterns
obtained by reflecting them vertically. The dots in the matrices are
essential: a row can continue all the way to the \emph{left boundary}
of the physical grid. Consequently the patterns apply to grids of
arbitrary finite width.

\subsection{How to read the matrices}\label{sec:extended-matrices}

A $0$ marks a position that is still possible but is \emph{not}
included in the next query, a $1$ marks a possible position included
in the query, and a dash indicates a position already excluded.
In every row, the first displayed symbol is repeated through
all columns further to the left, up to the actual left boundary.
In particular, a row beginning with $1$ queries a complete left
prefix; a row beginning with $0$ leaves that prefix unqueried; and
a row beginning with a dash has no possible positions there.
Every position to the right of the displayed matrix is excluded.
Only the intersection of each pattern with the physical $M\times N$
grid is used. If the pattern extends beyond the top, bottom, or
left boundary, those positions are simply absent. We retain the
\emph{virtual position} of the displayed pattern after any such
truncation; we never move it back inside the grid artificially.

For any one of the displayed matrices, one may also reverse the
order of its rows. This is its \emph{vertical reflection}. Reflecting
the entire pattern, including its repeated left prefixes, leaves
all arguments unchanged after exchanging ``above'' and ``below''.
Whenever a nonterminal answer is received, the target may make one
admissible move before the next query. If a terminal answer is
received, we stop \emph{immediately}, without allowing any
additional movement. Thus the accuracy is the number of possible
positions marked by the answer to the \emph{last} query.

\subsection{The initial staircase and nine query patterns}

The following four-row staircase will occur after an initial query
has received answer~$0$. It is \emph{not itself a query matrix}:
here its $1$-symbols merely display the \emph{shape of the remaining
candidate set}. Its upper row extends one column farther right
than its second row, and so on.

\[
\text{Starting shape}\quad\begingroup\setlength{\arraycolsep}{3.2pt}
\begin{pmatrix}
1 & 1 & 1 & \cdots & 1 & 1 & 1 & 1 & 1 & - \\
1 & 1 & 1 & \cdots & 1 & 1 & 1 & 1 & - & - \\
1 & 1 & 1 & \cdots & 1 & 1 & 1 & - & - & - \\
1 & 1 & 1 & \cdots & 1 & 1 & - & - & - & -
\end{pmatrix}\endgroup
\]

After one target movement, this shape expands into a six-row shape.
Matrix~$A$ below describes it and prescribes the next query.
In the nine matrices that follow, the symbols $0$ and $1$ have
their normal query meanings specified above.

\[
\text{Matrix }\mathsf{A}\quad\begingroup\setlength{\arraycolsep}{3.2pt}
\begin{pmatrix}
0 & 0 & 0 & \cdots & 0 & 0 & 1 & 1 & 1 & - \\
0 & 0 & 0 & \cdots & 0 & 0 & 0 & 1 & 1 & 1 \\
0 & 0 & 0 & \cdots & 0 & 0 & 0 & 1 & 1 & - \\
0 & 0 & 0 & \cdots & 0 & 0 & 0 & 0 & - & - \\
0 & 0 & 0 & \cdots & 0 & 0 & 0 & - & - & - \\
0 & 0 & 0 & \cdots & 0 & 0 & - & - & - & -
\end{pmatrix}\endgroup
\]

\[
\text{Matrix }\mathsf{B}\quad\begingroup\setlength{\arraycolsep}{3.2pt}
\begin{pmatrix}
0 & 0 & 0 & \cdots & 0 & 0 & - & - & - & - \\
0 & 0 & 0 & \cdots & 0 & 0 & 0 & - & - & - \\
0 & 0 & 0 & \cdots & 0 & 0 & 0 & 0 & - & - \\
0 & 0 & 0 & \cdots & 0 & 0 & 0 & 0 & - & - \\
1 & 1 & 1 & \cdots & 1 & 1 & 1 & 1 & 1 & - \\
1 & 1 & 1 & \cdots & 1 & 1 & 1 & 1 & - & - \\
1 & 1 & 1 & \cdots & 1 & 1 & 1 & - & - & - \\
1 & 1 & 1 & \cdots & 1 & 1 & - & - & - & -
\end{pmatrix}\endgroup
\]

\[
\text{Matrix }\mathsf{C}\quad\begingroup\setlength{\arraycolsep}{3.2pt}
\begin{pmatrix}
0 & 0 & 0 & \cdots & 0 & 0 & - & - & - & - \\
0 & 0 & 0 & \cdots & 0 & 0 & 0 & - & - & - \\
0 & 0 & 0 & \cdots & 0 & 0 & 0 & 1 & - & - \\
0 & 0 & 0 & \cdots & 0 & 0 & 0 & 1 & 1 & - \\
0 & 0 & 0 & \cdots & 0 & 0 & 0 & 1 & 1 & - \\
0 & 0 & 0 & \cdots & 0 & 1 & 1 & 1 & - & -
\end{pmatrix}\endgroup
\]

\[
\text{Matrix }\mathsf{D}\quad\begingroup\setlength{\arraycolsep}{3.2pt}
\begin{pmatrix}
0 & 0 & 0 & \cdots & 0 & 0 & - & - & - & - \\
0 & 0 & 0 & \cdots & 0 & 0 & 0 & - & - & - \\
0 & 0 & 0 & \cdots & 0 & 0 & 0 & 0 & - & - \\
0 & 0 & 0 & \cdots & 0 & 0 & 0 & 0 & - & - \\
1 & 1 & 1 & \cdots & 1 & 1 & 1 & 1 & - & - \\
1 & 1 & 1 & \cdots & 1 & 1 & 1 & 1 & - & - \\
1 & 1 & 1 & \cdots & 1 & 1 & 1 & - & - & - \\
1 & 1 & 1 & \cdots & 1 & - & - & - & - & -
\end{pmatrix}\endgroup
\]

\[
\text{Matrix }\mathsf{E}\quad\begingroup\setlength{\arraycolsep}{3.2pt}
\begin{pmatrix}
0 & 0 & 0 & \cdots & 0 & 0 & 1 & 1 & - & - \\
0 & 0 & 0 & \cdots & 0 & 0 & 0 & 1 & 1 & - \\
0 & 0 & 0 & \cdots & 0 & 0 & 0 & 1 & 1 & - \\
0 & 0 & 0 & \cdots & 0 & 0 & 0 & 1 & - & - \\
0 & 0 & 0 & \cdots & 0 & 0 & 1 & - & - & - \\
0 & 0 & 0 & \cdots & 0 & - & - & - & - & -
\end{pmatrix}\endgroup
\]

\[
\text{Matrix }\mathsf{F}\quad\begingroup\setlength{\arraycolsep}{3.2pt}
\begin{pmatrix}
0 & 0 & 0 & \cdots & 0 & 0 & - & - & - & - \\
0 & 0 & 0 & \cdots & 0 & 0 & 0 & - & - & - \\
0 & 0 & 0 & \cdots & 0 & 0 & 0 & 0 & - & - \\
0 & 0 & 0 & \cdots & 0 & 0 & 1 & 1 & - & - \\
1 & 1 & 1 & \cdots & 1 & 1 & 1 & 1 & - & - \\
1 & 1 & 1 & \cdots & 1 & 1 & 1 & - & - & - \\
1 & 1 & 1 & \cdots & 1 & 1 & - & - & - & - \\
1 & 1 & 1 & \cdots & 1 & - & - & - & - & -
\end{pmatrix}\endgroup
\]

\[
\text{Matrix }\mathsf{G}\quad\begingroup\setlength{\arraycolsep}{3.2pt}
\begin{pmatrix}
- & - & - & \cdots & - & - & 1 & 1 & - & - \\
0 & 0 & 0 & \cdots & 0 & 0 & 1 & 1 & 1 & - \\
0 & 0 & 0 & \cdots & 0 & 0 & 0 & 1 & 1 & - \\
0 & 0 & 0 & \cdots & 0 & 0 & 0 & 1 & - & - \\
0 & 0 & 0 & \cdots & 0 & 0 & 0 & - & - & - \\
0 & 0 & 0 & \cdots & 0 & 0 & - & - & - & - \\
0 & 0 & 0 & \cdots & 0 & - & - & - & - & -
\end{pmatrix}\endgroup
\]

\[
\text{Matrix }\mathsf{H}\quad\begingroup\setlength{\arraycolsep}{3.2pt}
\begin{pmatrix}
0 & 0 & 0 & \cdots & 0 & 1 & - & - & - & - \\
0 & 0 & 0 & \cdots & 0 & 0 & 1 & - & - & - \\
0 & 0 & 0 & \cdots & 0 & 0 & 0 & 1 & - & - \\
0 & 0 & 0 & \cdots & 0 & 0 & 0 & 1 & 1 & - \\
0 & 0 & 0 & \cdots & 0 & 0 & 1 & 1 & - & - \\
0 & 0 & 0 & \cdots & 0 & 1 & - & - & - & -
\end{pmatrix}\endgroup
\]

\[
\text{Matrix }\mathsf{I}\quad\begingroup\setlength{\arraycolsep}{3.2pt}
\begin{pmatrix}
0 & 0 & 0 & \cdots & 0 & - & - & - & - & - \\
0 & 0 & 0 & \cdots & 0 & 0 & - & - & - & - \\
0 & 0 & 0 & \cdots & 0 & 0 & 0 & - & - & - \\
0 & 0 & 0 & \cdots & 0 & 0 & 0 & 0 & - & - \\
1 & 1 & 1 & \cdots & 1 & 1 & 1 & 1 & - & - \\
1 & 1 & 1 & \cdots & 1 & 1 & 1 & - & - & - \\
1 & 1 & 1 & \cdots & 1 & 1 & - & - & - & - \\
1 & 1 & 1 & \cdots & 1 & - & - & - & - & -
\end{pmatrix}\endgroup
\]

The query sets marked by $1$ in matrices $A,C,E,G,H$ have exactly
eight vertices in their untruncated forms: the respective row-wise
counts are
\[
 A:\ 3+3+2,\qquad
 C:\ 1+2+2+3,\qquad
 E:\ 2+2+2+1+1,\qquad
 G:\ 2+3+2+1,\qquad
 H:\ 1+1+1+2+2+1.
\]
These queries contain no repeated left prefix of $1$-symbols.
Hence, whenever the answer is $1$ in any of these five matrices,
the strategy terminates with at most eight possible positions.
In the remaining matrices $B,D,F,I$, the queries may contain an
arbitrarily long left prefix, so their positive answers generally
require another query.

\subsection{The theorem and complete proof}\label{sec:nine-transitions}

\begin{theorem}[Accuracy eight on every finite rectangular grid]
\label{thm:accuracy-eight}
For every pair of positive integers $M,N$, there exists an adaptive
search strategy such that, after finitely many queries, at most eight
vertices remain possible \emph{immediately after its last answer}.
In particular,
\[
 \boxed{s^*(M,N)\leq8.}
\]
\end{theorem}

\begin{proof}
We describe the initial reduction, verify the complete decision
procedure for matrices $A,\ldots,I$, and prove that it cannot
continue indefinitely on a finite grid.

\smallskip
\noindent\textbf{Step 1: Reaching the starting staircase.}
If there are at least two rows, Proposition~\ref{prop:two-row-reduction}
first restricts the candidates to two adjacent complete rows. After
the next possible target movement, candidates can lie in at most
four consecutive complete rows. We refer to these rows, including
any virtual rows outside the actual boundary, from top to bottom.

We now use a single \emph{staircase query}. In the highest of
these four rows we query no positions. In the next row we query
only the rightmost position, in the third row the two rightmost
positions, and in the lowest row the three rightmost positions.
Positions outside the physical grid are, of course, ignored.
The test therefore contains at most
\[
                         0+1+2+3=6
\]
positions. If the answer is $1$, the possible positions are
contained in this test and we stop immediately with accuracy
at most six.

Suppose instead that the answer is $0$. In each of the four
rows, precisely the queried right-hand positions are eliminated.
Accordingly, the surviving positions are contained in the
\emph{starting staircase}: the highest row reaches the original
right boundary, the next row ends one column earlier, and the
remaining two rows end two and three columns earlier,
respectively. These are the four rows of the displayed starting
shape. Near a boundary, some rows or right-hand portions may be
absent; we always use the intersection of this shape with the
physical grid, without changing its placement.

Before the next query, the target may make one move. The starting
staircase can then expand by one row upwards and one row
downwards. To determine the new shape, look only at the
\emph{rightmost possible position} in each row. For the six
successive rows, from top to bottom, those right ends lie
\[
\begin{gathered}
\text{at the original right boundary},\quad
\text{one column beyond it},\quad
\text{at the original right boundary},\\
\text{one column to its left},\quad
\text{two columns to its left},\quad
\text{three columns to its left}.
\end{gathered}
\]
The second right end is necessarily cut off by the physical
right boundary. All surviving rows extend to the left boundary,
where present. Thus their possible positions are contained
\emph{exactly in the support} of matrix $A$ after physical
truncation. Applying matrix $A$ as the next query is therefore
valid also for short or narrow grids. The remaining proof treats
all possible answers to $A,\ldots,I$ and their reflections.

\smallskip
\noindent\textbf{Step 2: All answers and their next states.}
For clarity, every shift stated below describes the \emph{top
row of the next full virtual matrix compared with the top row of
the current full virtual matrix}. A horizontally shifted matrix
has all of its displayed columns moved together. The table
gives every possible continuation. ``Stop'' means stop
\emph{before} any further target movement.

\begin{center}
\small
\renewcommand{\arraystretch}{1.27}
\begin{tabular}{@{}c p{0.38\linewidth} p{0.38\linewidth}@{}}
\toprule
Query & Answer $0$ & Answer $1$\\
\midrule
$A$ & $B$, beginning one row above & Stop, at most $8$ positions\\
$B$ & $C$, beginning one row above & $A$, beginning three rows below\\
$C$ & $D$, beginning one row above & Stop, at most $8$ positions\\
$D$ & $C$, beginning one row above & $E$, beginning three rows below\\
$E$ & $F$, beginning one row above & Stop, at most $8$ positions\\
$F$ & $H$, beginning one row above & $G$, beginning two rows below\\
$G$ & $F$, with the same top row & Stop, at most $8$ positions\\
$H$ & $I$, beginning one row above & Stop, at most $8$ positions\\
$I$ & Reflected $A$, beginning one row above, and shifted one column left
    & $A$, beginning three rows below, and shifted one column left\\
\bottomrule
\end{tabular}
\end{center}

Here is the geometric reading of the first two phases. We
spell out both possible answers, because the points of departure
from the two cycles are crucial to the argument.

\medskip
\noindent\emph{First phase: matrices $A$ and $B$.}
The eight $1$-positions in $A$ lie in the upper three rows,
close to the right-hand end of the current staircase. If the
answer to $A$ is $1$, the target must occupy one of those eight
positions, so the search ends \emph{immediately}; no further
movement is included in the accuracy. If the answer is $0$,
only the $0$-positions remain possible. They occupy six
left-extending rows. Allowing one movement enlarges these
rows to the eight-row pattern $B$, whose top row is one row
\emph{above} the top row of the preceding $A$.

Matrix $B$ tests the four \emph{lower} rows, including their
whole possible left-hand portions. If its answer is $1$, those
four rows are retained. After one movement their upper and
lower edges expand to six rows, forming another copy of $A$.
This new $A$ starts three rows \emph{below} the top row of $B$;
because $B$ started one row above the previous $A$, the net
effect of the two tests is that $A$ has moved \emph{two rows
down}. The repeated answer sequence $A:0$, $B:1$ therefore
creates a downward-moving cycle, not a stationary loop.

If, on the other hand, the answer to $B$ is $0$, only the four
\emph{upper} rows remain. After the target's movement these
expand to six rows and form matrix $C$, beginning one row
\emph{above} $B$. Thus the very answer which breaks the
$A\to B\to A$ cycle also starts the next phase. The lower
boundary makes indefinite repetition impossible: eventually
$B$ can have no possible $1$-position in its lower four rows,
so another answer $1$ is impossible and the feasible response
is $0$ (unless an earlier $A$ has already ended the search).
This also handles the last, possibly truncated, copy of $B$:
one does \emph{not} assume that a full pattern fits inside the
grid before making the final test.

\medskip
\noindent\emph{Second phase: matrices $C$ and $D$.}
Matrix $C$ places its eight $1$-positions near the right-hand
end of four of its six rows. A positive answer is therefore
terminal, again with accuracy at most eight measured \emph{at
that answer}. A negative answer retains the left-extending
$0$-positions. After one movement they become the eight-row
pattern $D$, beginning one row \emph{above} $C$.

Matrix $D$ now queries its four \emph{lower} rows. If its
answer is $0$, only the four \emph{upper} rows survive. Their
one-step enlargement is another copy of $C$, which begins one
row above $D$. Together with the preceding shift from $C$ to
$D$, this moves the new $C$ \emph{two rows up}. In other words,
$C:0$, $D:0$ is an upward-moving cycle.

If instead the answer to $D$ is $1$, its four lower rows are
retained. Their one-step enlargement forms the six-row matrix
$E$, beginning three rows \emph{below} $D$. This is the exit
from the second cycle. At the upper boundary, the rows of $D$
that carry $0$-positions eventually lie entirely outside the
physical grid. In that last boundary configuration, answer $0$
cannot occur; a feasible nonterminal history must follow
$D:1$ and proceed to $E$, unless $C:1$ has already terminated
the search. This argument applies equally to a truncated copy
of $D$.

\medskip
\noindent\emph{Third phase: matrices $E$, $F$, and $G$.}
The eight queried positions of $E$ are located near the right
ends of its rows. Hence the answer $1$ ends the search
immediately. If the answer is $0$, we retain the left-extending
parts of the six rows and let the target move once. This gives
the eight-row pattern $F$, whose top row lies one row
\emph{above} the top row of $E$.

Matrix $F$ is different from $E$: its last four rows contain
long queried prefixes extending to the left boundary, while
its first three rows contain only unqueried positions. The
fourth row is divided into an unqueried left part and a short
queried right part. Therefore neither answer to $F$ need be
terminal.

If $F$ gives answer $1$, we retain the queried part of the
fourth row together with the queried lower rows. After one
movement these positions form the seven-row pattern $G$.
The top row of $G$ is \emph{two rows below} the top row of $F$.
Notice that the top row of $G$ consists of just two isolated
possible positions, not an entire prefix extending to the
left boundary. All eight $1$-positions of $G$ lie in a small
region near the right end of the current pattern. Thus
$G:1$ ends the search with at most eight positions.

If instead $G$ gives answer $0$, the eight queried positions
are excluded and the left-extending unqueried portions of
its lower six rows remain. Their enlargement by one movement
reproduces $F$, with its top row in the \emph{same} place as
the top row of $G$. As $G$ started two rows below the previous
$F$, every full sequence $F:1$, $G:0$ moves the next copy
of $F$ \emph{two rows down}. The cycle therefore cannot be
repeated indefinitely on a finite-height grid. Its exit is
$F:0$: in this case only the unqueried upper portion of $F$
remains, and after one movement we obtain $H$, beginning one
row above $F$.

\medskip
\noindent\emph{Fourth phase: matrices $H$ and $I$.}
The eight queried positions of $H$ form a narrow zigzag on
the right side of its six rows. A positive answer is terminal.
A negative answer keeps the unqueried left prefixes in the
six rows. Allowing one movement produces the eight-row
pattern $I$, beginning one row above $H$.

Matrix $I$ splits its possibilities into two groups: the four
upper rows are unqueried, while the four lower rows are
queried, in both cases extending to the left boundary.
If the answer to $I$ is $1$, we retain the lower four rows;
after the next movement we obtain the \emph{unreflected}
pattern $A$, whose top row is three rows below that of $I$.
If the answer is $0$, we retain the upper four rows; after the
movement we obtain the \emph{vertically reflected} pattern
$A$, whose top row is one row above that of $I$.

The decisive fact is common to \emph{both} answers: the new
copy of $A$ is shifted \emph{one column to the left}. A
reflection changes only which way is called ``up'' or
``down''; it does not undo that leftward shift. The procedure
therefore restarts at $A$, possibly reflected, but with a
strictly smaller horizontal placement. The left boundary of
the finite grid will eventually prevent an unlimited number
of such restarts.

\smallskip
\noindent\textbf{Step 3: Verification of the matrix transitions.}
We verify the transition table first for the complete virtual
patterns. We then explain why the same transitions remain valid
when parts of a pattern lie outside the physical grid.
For an answer $0$, erase all $1$-positions of the current matrix;
for an answer $1$, erase all $0$-positions. Before the next query,
allow each retained position to stay or move one unit horizontally
or vertically. Each next row therefore consists of the retained
positions in the row itself, their immediate left and right
neighbors, and the retained positions directly above and below.
Repeat the leading symbols of all rows to the left before taking
this neighborhood. The resulting candidate set is contained
in the matrix shown in the corresponding entry of the table.

We can verify the two displacement cycles directly from the
matrices. For $A:0$, the six surviving left prefixes have their
right-hand ends arranged in the staircase displayed in $A$.
After one movement they occupy eight consecutive rows, exactly
as in $B$ placed one row above $A$. For $B:1$, the four
queried lower prefixes remain. One movement creates six rows,
with the longest rightward extent in the second of these rows,
exactly as in $A$ placed three rows below $B$. For $B:0$, the
four unqueried upper prefixes remain; one movement creates
six rows whose right-hand ends agree with matrix $C$, one row
above $B$.

Likewise, after $C:0$, expanding the six left-extending rows
produces $D$ placed one row above $C$. After $D:0$, expanding
its four unqueried upper rows produces another $C$, one row
above $D$; after $D:1$, expanding its four queried lower rows
produces $E$, three rows below $D$. These equalities make
both the direction of each shift and the exit from each cycle
explicit. Near a physical boundary the same calculations
produce subsets of these full virtual shapes, which suffices
for the strategy.

The remaining transitions can be checked just as explicitly.
For $E:0$, the unqueried portions of the six successive rows
extend to the left and end, relative to the first displayed
column, at offsets
\[
                        1,\ 2,\ 2,\ 2,\ 1,\ 0.
\]
After one allowed movement, the eight rows have right-hand
ends at offsets
\[
                    1,\ 2,\ 3,\ 3,\ 3,\ 2,\ 1,\ 0.
\]
These are exactly the ends of the possible-position rows in
$F$, starting one row above $E$.

For $F:1$, the surviving queried positions lie in the fourth
through eighth rows. After one movement, their support is
precisely the seven-row pattern $G$, two rows below $F$.
In particular, the \emph{first} row of $G$ consists of only
two possible positions (the two $1$-symbols in that row):
it has no left-extending prefix. For $G:0$, the queried
positions are removed; the remaining six left prefixes have
right-hand ends at offsets
\[
                        1,\ 2,\ 2,\ 2,\ 1,\ 0.
\]
Their one-step enlargement has ends
\[
                    1,\ 2,\ 3,\ 3,\ 3,\ 2,\ 1,\ 0,
\]
which reproduce $F$ with the same top row as $G$. This
proves both directions of the $F,G$ cycle.

For $F:0$, the surviving upper four left prefixes, including
the unqueried portion of the fourth row, expand to the six
rows of $H$. Their right-hand ends are, from top to bottom,
\[
                         1,\ 2,\ 3,\ 4,\ 3,\ 1.
\]
For $H:0$, the six remaining unqueried prefixes expand to
eight rows with right-hand ends
\[
                    0,\ 1,\ 2,\ 3,\ 3,\ 2,\ 1,\ 0.
\]
This is precisely the possible-position support of $I$,
placed one row above $H$. Thus neither transition relies on
an unverified picture or on an assumption about the length
of the left prefixes.

It is particularly important to check \emph{both} answers in $I$.
For answer $0$, retain the four upper rows of $I$ and discard its
four lower rows. Count columns from the first displayed column of
$I$, numbered $0,1,2,\ldots$. The rightmost surviving positions
in those four rows have indices $0,1,2,3$. After one movement,
the new six successive rows have right endpoints
\[
                 0,\ 1,\ 2,\ 3,\ 4,\ 3.
\]
These are exactly the row endpoints of a \emph{vertically reflected}
$A$ whose displayed columns have moved one position left. For
answer $1$, retain instead the four lower rows, with right
endpoints $3,2,1,0$. Their neighborhood has six successive
right endpoints
\[
                 3,\ 4,\ 3,\ 2,\ 1,\ 0,
\]
which are exactly those of the unreflected $A$ after shifting
its displayed columns one position left. This explicit
calculation proves the essential leftward move on \emph{either}
answer in $I$. The identities apply to the complete left-extended
matrices, not only to the six columns printed on the page.

All other entries of the transition table follow by this same
row-by-row neighborhood check. More explicitly, the thirteen
nonterminal checks are
\[
\begin{gathered}
 A\xrightarrow{0}B,\quad
 B\xrightarrow{0}C,\quad B\xrightarrow{1}A,\\
 C\xrightarrow{0}D,\quad
 D\xrightarrow{0}C,\quad D\xrightarrow{1}E,\\
 E\xrightarrow{0}F,\quad
 F\xrightarrow{0}H,\quad F\xrightarrow{1}G,\\
 G\xrightarrow{0}F,\quad
 H\xrightarrow{0}I,\quad
 I\xrightarrow{0}\overline A,\quad I\xrightarrow{1}A,
\end{gathered}
\]
with exactly the shifts already specified in words. All thirteen checks give \emph{equality} for the complete
left-extended patterns, before they are restricted to the physical
grid. In each case, the rightmost occupied site of every nonempty
row and the distinction between a left prefix and a finite interval
uniquely determine the expanded set.

The use of virtual patterns at a boundary is justified by a simple
containment. Let $V$ be the vertex set of the physical finite grid,
and let $S$ be any virtual set of positions in the integer lattice.
Then
\[
  \Gamma_{G_{M,N}}(S\cap V)
  \subseteq \Gamma_{\mathbb Z^2}(S)\cap V.
\]
Indeed, each move permitted within the grid is also a permitted
move in the integer lattice; clipping away virtual positions cannot
create additional possible positions. Applying this containment
inductively to each of the thirteen exact virtual transitions shows
that the \emph{actual} candidate set is always contained in the
stated next matrix restricted to the physical grid. This is why
truncated copies require no separate transition rules, even when
the initial pattern overhangs the top, bottom, or right boundary.
Vertical reflection preserves the same containment and proves all
reflected cases.

\smallskip
\noindent\textbf{Step 4: Why the cycles must end, including the last
possible answer at a boundary.}
There are only three possible cycles that do not use $I$:
\[
 A\longrightarrow B\longrightarrow A,\qquad
 C\longrightarrow D\longrightarrow C,\qquad
 F\longrightarrow G\longrightarrow F.
\]
Each complete repetition of the first and third cycle starts
its next copy two rows \emph{lower}; each repetition of the
second starts its next copy two rows \emph{higher}. For the
vertically reflected patterns the directions are reversed.
No repetition increases the number of virtual rows in any of
these patterns: every pattern occupies at most eight consecutive
rows. Therefore, on a grid with finitely many rows, none of
these cycles can be followed indefinitely by a nonempty
candidate set.

The \emph{last feasible step at a boundary} deserves special
attention. Consider the $A,B$ cycle. Every full repetition
moves the virtual pattern two rows downward. Eventually the
four lower rows queried in $B$ have no intersection with the
physical grid; at that moment $B:1$ is impossible. If the
target is still among the candidates, its answer to this last
$B$ must be $0$, which leads to $C$. Before this point, a
partially truncated $B$ is handled by exactly the same query
rule, so no extra boundary test is needed.

For $C,D$ the situation is the vertical counterpart: each
repetition moves $C$ two rows upward. Eventually the upper
four unqueried rows of $D$ are entirely above the physical
grid, making $D:0$ impossible. If the search has not already
stopped at $C:1$, the last feasible answer in $D$ is $1$ and
the strategy moves to $E$. Again, the matrices are always
intersected with the physical grid, even during the final
truncated cycle.

Finally, in the $F,G$ cycle every full repetition shifts
$F$ down by two rows. The $1$-positions of $F$ occur only
in its fourth through eighth rows. Once that queried region
has moved entirely below the physical grid, answer $1$ in
$F$ is impossible. A feasible continuation must then use
$F:0$ and proceed to $H$, unless an earlier $G:1$ already
terminated the search. Partially truncated copies are handled
by intersecting the same patterns with the physical grid,
including the last such $F$ query. Vertical reflection
exchanges upper and lower boundaries without changing the
argument. All three cycles therefore have only finitely
many \emph{feasible} repetitions; no additional ``last
question'' or extra movement is silently assumed.

It remains to exclude infinitely many returns through $I$.
The first copy of $A$ is anchored at the physical right boundary:
its rightmost virtual position lies only one column beyond that
boundary, and the physical candidate set is obtained by clipping
the virtual support. None of the transitions within the phases
$A,B$, $C,D$, or $E,F,G,H$ changes the horizontal placement.
By contrast, \emph{each} departure from $I$ or its reflected
version shifts the entire next copy of $A$ one column to the left.
This shift is the same for both answers to $I$.
After sufficiently many visits, the rightmost position of the
\emph{virtual} pattern would lie strictly to the left of the
physical grid. Its intersection with the grid would then be empty,
contradicting feasibility of the answer history. Hence only
finitely many visits to $I$ are possible. This conclusion does
not assume that any particular answer must occur at the left
boundary: it follows from the strict leftward progress of both
possible answers.

Suppose for contradiction that a feasible execution contains
infinitely many tests. Since only finitely many visits to $I$
are possible, after the last such visit only the other eight
states, with fixed reflection orientation, can occur.
Apart from the three two-state cycles listed above, these
states are traversed only in the forward order of the phases.
An infinite execution must therefore repeat one of the three
cycles forever, contradicting its strict vertical drift on
a finite-height grid. Hence \emph{every} feasible execution
terminates after finitely many tests. Since the strategy is
binary branching, the absence of an infinite feasible branch
also implies that its feasible decision tree is finite.

\smallskip
\noindent\textbf{Step 5: Accuracy at termination.}
The only terminal transitions of the nine-pattern strategy
are the positive answers in $A,C,E,G,H$, and each corresponding
query contains at most eight positions. The initial staircase
may also stop at a positive answer, with at most six positions.
No target movement occurs after these final answers. Therefore
the final candidate set always contains at most eight vertices.
If the grid has only one row, transpose the roles of rows and
columns and apply the already proved case with at least two
rows; the single-vertex grid is immediate. This proves
$s^*(M,N)\leq8$ for all finite $M,N$.
\end{proof}

\section{An adaptive diagonal sweep}\label{sec:diagonal-sweep}

The sweep tests two adjacent rows at a time. Within each pair, the
queried part of the upper row shrinks from the left, while the
queried part of the lower row grows from the left. The strategy uses
each answer to decide whether to stop or continue. A positive answer
terminates the search; after $N$ consecutive negative answers within
a row pair, the sweep advances by one row.

\begin{theorem}[Dimension-dependent upper bound]
\label{thm:diagonal-sweep}
For all $M,N\geq1$,
\begin{equation}\label{eq:diagonal-sweep-bound}
 s^*(M,N)\leq\min\{M+1,N+1\}.
\end{equation}
\end{theorem}
\begin{proof}
First suppose $M\geq2$. For each $r\in[M-1]$ and $j\in[N]$,
define the query
\begin{equation}\label{eq:diagonal-sweep-query}
 T_{r,j}=\bigl(\{r\}\times\{j,\ldots,N\}\bigr)
 \cup\bigl(\{r+1\}\times\{1,\ldots,j\}\bigr).
\end{equation}
Thus $|T_{r,j}|=N+1$. Along the branch on which all preceding
answers were negative, use these queries in lexicographic order,
$T_{1,1},\ldots,T_{1,N},T_{2,1},\ldots,T_{M-1,N}$.
After each answer, stop if it is positive and otherwise continue
with the next query. This answer-dependent stopping rule is part
of the adaptive strategy. A positive answer leaves at most $N+1$
possible positions, since no movement follows the terminal test.

We show that a sequence of negative answers cannot continue past
the end of the schedule. For any pair $(r,j)$ at which a negative
answer occurs, let $D_{r,j}$ be the candidate set immediately after
that answer. We claim the invariant
\begin{equation}\label{eq:diagonal-sweep-invariant}
 D_{r,j}\subseteq
 \bigl(\{r+1\}\times\{j+1,\ldots,N\}\bigr)
 \cup\bigl(\{r+2,\ldots,M\}\times[N]\bigr),
\end{equation}
where an interval with its lower endpoint greater than its upper
endpoint denotes the empty set.

At the beginning of phase $r=1$, all positions are possible before
$T_{1,1}$. This query tests the complete first row and position
$(2,1)$. A negative answer therefore leaves only positions in the
second row with column at least $2$, or in lower rows, proving
\eqref{eq:diagonal-sweep-invariant} for $(r,j)=(1,1)$.

For the induction within a phase, suppose that
\eqref{eq:diagonal-sweep-invariant} holds after a negative answer
to $T_{r,j-1}$, with $2\leq j\leq N$. Before the next test the
target may move by at most one grid edge. Hence no position above
row $r$ can become possible, and every possible position in row
$r$ must have column at least $j$: it can only have been reached
from row $r+1$ at the same column. The query $T_{r,j}$ includes
all these positions in row $r$ and columns $1,\ldots,j$ in row
$r+1$. A negative answer thus establishes
\eqref{eq:diagonal-sweep-invariant} for $(r,j)$.

Finally, if all tests in phase $r$ have negative answers, then
\eqref{eq:diagonal-sweep-invariant} at $j=N$ yields
\[
 D_{r,N}\subseteq\{r+2,\ldots,M\}\times[N].
\]
After the next movement, the target can occupy only rows
$r+1,\ldots,M$. Therefore the next phase starts under the same
condition as the first: $T_{r+1,1}$ tests its entire upper row
and the first position of the following row. Its negative branch
satisfies the invariant. This completes the induction across
phases.

If every scheduled answer were negative, then
\eqref{eq:diagonal-sweep-invariant} for $(r,j)=(M-1,N)$ would
force $D_{M-1,N}=\varnothing$, contradicting feasibility.
Consequently a positive answer occurs before the finite schedule
is exhausted, and every terminal candidate set has size at most
$N+1$.

Interchanging rows and columns gives the bound with $M+1$.
If $M,N\geq2$, taking the better of these two accuracy bounds
proves \eqref{eq:diagonal-sweep-bound}. If exactly one dimension
is $1$, apply the sweep in the orientation with at least two
rows to obtain accuracy $2$.
For the $1\times1$ grid, one test leaves a singleton.
\end{proof}

\begin{remark}
The stopping rule is essential: the search ends at the first positive
answer, and the accuracy bound applies immediately after that answer,
without a further movement of the target. The theorem concerns the
adaptive accuracy $s^*(M,N)$; it does not assert the same bound for
a strategy whose termination is fixed before any answers are observed.
\end{remark}

\section{Search on a Finite Cylindrical Grid}\label{sec:cylindrical-grid}

We now consider a finite grid with periodic boundary conditions in
one direction. In contrast to a rectangular grid, the first and last
columns are adjacent. In contrast to a torus, the first and last
rows remain genuine boundaries. The number of rows and the number of
columns are both finite.

\begin{definition}[Cylindrical grid]\label{def:cylindrical-grid}
For integers $M\geq 1$ and $N\geq 3$, the \emph{cylindrical grid} is
\[
  G^{\mathrm{cyl}}_{M,N}=P_M\square C_N,
  \qquad V^{\mathrm{cyl}}_{M,N}=[M]\times[N].
\]
Two distinct vertices are adjacent if their positions differ by one
row in the vertical direction or by one column in the horizontal
direction, with the horizontal coordinate understood cyclically.
In particular, a target at any position in column~$1$ may move
horizontally to the corresponding position in column~$N$, and vice
versa. There is no corresponding move between the top and bottom
rows. As in the rectangular model, remaining at the current vertex
is permitted between consecutive queries, whereas diagonal moves
are not.
\end{definition}

\begin{definition}[Queries, candidate sets, and terminal accuracy]
\label{def:cylindrical-accuracy}
Initially all vertices are possible. Before each query, the target
may stay where it is or move to an adjacent vertex of the
cylindrical grid. A query is an arbitrary subset of vertices, and
its binary answer indicates whether the current target position
belongs to that subset. Subsequent queries may be chosen adaptively
from the answers already received.

Let $\Gamma_{\mathrm{cyl}}(S)$ denote the set consisting of $S$
and all vertices that can be reached from $S$ by one permitted move.
If $D_{t-1}$ is the candidate set immediately after the previous
answer, then the candidates immediately before the next query are
\[
   C_t=\Gamma_{\mathrm{cyl}}(D_{t-1}).
\]
For a query $T_t\subseteq V^{\mathrm{cyl}}_{M,N}$ with answer
$y_t\in\{0,1\}$, the updated candidate set is
\[
 D_t=\begin{cases}
 C_t\cap T_t, & y_t=1,\\
 C_t\setminus T_t, & y_t=0.
 \end{cases}
\]
Here $D_0=V^{\mathrm{cyl}}_{M,N}$; hence the first movement does
not enlarge the initial uncertainty. A strategy must terminate after
finitely many queries for every feasible sequence of answers.

The \emph{accuracy} of a strategy is the largest number of candidate
positions remaining \emph{immediately after the last answer}.
No additional target movement is counted after termination.
A strategy is called \emph{$s$-successful} if it has this
termination property and always leaves at most $s$ candidates.
The optimal cylindrical accuracy is
\[
 s^*_{\mathrm{cyl}}(M,N)
 =\min\bigl\{s\geq1:\text{an $s$-successful strategy exists}\bigr\}.
\]
\end{definition}

\begin{definition}[Reading the local query matrices]
\label{def:cylindrical-matrix-convention}
The matrices used to specify a strategy describe candidate positions
and the current query, rather than the geometric boundaries of the
grid. An entry $0$ denotes a possible target position omitted from
the query, an entry $1$ denotes a possible position included in the
query, and a dash $-$ denotes a position excluded by the preceding
answers. The printed matrices may show only a portion of the cyclic
sequence of columns: further possible positions, represented by
zeros, may lie to the left of the printed portion. In particular,
the first displayed column need not coincide with column~$1$ of
the cylinder. All positions and moves must nevertheless be
interpreted on the entire cylindrical grid, including adjacency
across the periodic identification of the first and last columns.
\end{definition}

\begin{theorem}[Proposed accuracy-nine bound for cylindrical grids]
\label{thm:cylinder-nine}
For every finite cylindrical grid with $M\geq1$ rows and
$N\geq3$ cyclically arranged columns, the proposed bound is
\[
 \boxed{\displaystyle s^*_{\mathrm{cyl}}(M,N)\leq9.}
\]
\end{theorem}
\begin{proof}[Proposed nine-position strategy and the remaining gap]
We describe the intended strategy on the cylindrical grid
$P_M\square C_N$ and isolate the point at which its correctness
still requires an additional argument. As in the definition, a $0$
represents a \emph{possible} target position outside the query, a
$1$ a possible position inside the query, and a dash an impossible
position. In particular, the leading zeros are \emph{fixed entries
of the current candidate configuration}. They are neither a
physical left boundary nor freely extendable padding. If the
actual candidate set contains further positions to the left of
the printed matrix, those positions must be specified by the
preceding answers and included in every subsequent calculation;
no additional zero entries may be inserted merely to make a
proposed transition work. Column indices are always read modulo
$N$.

\smallskip
\noindent\textbf{The proposed succession of query patterns.}
The remaining queries in the strategy are described by the nine
local patterns $A_9,\ldots,I_9$ in the table below. The
ellipsis in a printed row represents only columns that are
\emph{actually present} in that configuration; it does not
indicate an arbitrary or adjustable number of leading zeros.
The rows and columns of a pattern may be translated, and an entire
pattern may be reflected vertically. The reflected pattern has
exactly the same meaning as the original one.
The words ``above'', ``below'', and ``left'' in the table refer
to the proposed shift of the \emph{whole candidate pattern}
after the answer and the ensuing movement; they do not introduce
a new boundary on the cylinder.

\begin{center}
\small
\renewcommand{\arraystretch}{1.14}
\begin{longtable}{@{}c p{0.34\linewidth} p{0.20\linewidth} p{0.20\linewidth}@{}}
\caption{Proposed nine-position query patterns. The indicated successors
are the intended rectangular-grid transitions; their validity on a
cylinder remains to be checked.}\label{tab:proof-cylinder-nine}\\
\toprule
State & Candidate/query pattern & Answer $0$ & Answer $1$\\
\midrule\endfirsthead
\toprule
State & Candidate/query pattern & Answer $0$ & Answer $1$\\
\midrule\endhead
\bottomrule\endfoot
$A_9$ & \shortstack[l]{\texttt{00...00111-}\\\texttt{00...000111}\\\texttt{00...00011-}\\\texttt{00...0001--}\\\texttt{00...000---}\\\texttt{00...00----}} & $B_9$ (one row above) & stop ($\leq9$)\\[4pt]
$B_9$ & \shortstack[l]{\texttt{00...00----}\\\texttt{00...000---}\\\texttt{00...0000--}\\\texttt{00...0000--}\\\texttt{11...1111--}\\\texttt{11...1111--}\\\texttt{11...111---}\\\texttt{11...11----}} & $C_9$ (one row above) & reflected $C_9$\\[4pt]
$C_9$ & \shortstack[l]{\texttt{00...01----}\\\texttt{00...000---}\\\texttt{00...0001--}\\\texttt{00...00011-}\\\texttt{00...00011-}\\\texttt{00...0111--}} & $D_9$ (one row above) & stop ($\leq9$)\\[4pt]
$D_9$ & \shortstack[l]{\texttt{00...0-----}\\\texttt{00...000---}\\\texttt{00...0000--}\\\texttt{00...0000--}\\\texttt{11...1111--}\\\texttt{11...1111--}\\\texttt{11...111---}\\\texttt{11...1-----}} & $E_9$ (one row above) & reflected $E_9$\\[4pt]
$E_9$ & \shortstack[l]{\texttt{00...0-----}\\\texttt{00...001---}\\\texttt{00...0001--}\\\texttt{00...00011-}\\\texttt{00...00011-}\\\texttt{00...0111--}} & $F_9$ (one row above) & stop ($\leq9$)\\[4pt]
$F_9$ & \shortstack[l]{\texttt{00...0-----}\\\texttt{00...00----}\\\texttt{00...000---}\\\texttt{00...0000--}\\\texttt{11...1101--}\\\texttt{11...1111--}\\\texttt{11...111---}\\\texttt{11...1-----}} & $H_9$ (one row above) & $G_9$ (three rows below)\\[4pt]
$G_9$ & \shortstack[l]{\texttt{00...01-1--}\\\texttt{00...00111-}\\\texttt{00...00011-}\\\texttt{00...0001--}\\\texttt{00...001---}\\\texttt{00...0-----}} & $I_9$ (one row above) & stop ($\leq9$)\\[4pt]
$H_9$ & \shortstack[l]{\texttt{00...0-----}\\\texttt{00...00----}\\\texttt{00...000---}\\\texttt{00...0001--}\\\texttt{00...00111-}\\\texttt{00...0111--}\\\texttt{---...---1---}} & $I_9$ (one row above) & stop ($\leq8$)\\[4pt]
$I_9$ & \shortstack[l]{\texttt{00...0-----}\\\texttt{00...00----}\\\texttt{00...000---}\\\texttt{00...0000--}\\\texttt{11...1111--}\\\texttt{11...111---}\\\texttt{11...11----}\\\texttt{11...1-----}} & reflected $A_9$ (one column left) & $A_9$ (one column left)\\
\end{longtable}
\end{center}

The only answers at which the proposed scheme stops are the
positive answers in $A_9,C_9,E_9,G_9,H_9$. Counting their
entries equal to $1$ gives respectively $9,9,9,9,8$.
Consequently, \emph{provided the claimed transitions are valid
and all feasible answer histories terminate}, every terminal
answer has the required accuracy. Other answers lead to the
successor state in the table, including the indicated vertical
reflections. The horizontal shifts occur only in the two
successors of $I_9$.

\end{proof}
% TODO (joint revision): Insert a complete proof of
% Theorem~\ref{thm:cylinder-nine} here after checking all cyclic
% candidate-set transitions, reflected cases, and termination.
% The theorem is at present a proposed assertion, NOT yet proved.

\section{Conclusion and Open Problems}\label{sec:conclusion}

We proved that accuracy eight is attainable by an adaptive search
strategy on every finite rectangular grid, regardless of its
dimensions. The proof is constructive and uses only nine extended
matrix templates and their vertical reflections. Both answers in a
distinguished state advance the search towards the left boundary,
while cycles that avoid this state move strictly towards a vertical
boundary. These two progress mechanisms establish termination
without a column-insertion induction.

We complemented the universal construction with the
dimension-dependent bound
\[
 s^*(M,N)\leq\min\{8,M+1,N+1\}.
\]
% Together with our lower bounds, the elementary diagonal sweep
% establishes the exact accuracies
% \[
%  s^*(3,N)=4\quad(N\geq3),\qquad
%  s^*(4,N)=5\quad(N\geq4).
% \]
% The geometric converse gives $s^*(M,N)\geq5$ for $M,N\geq4$.

For finite cylindrical grids $P_M\square C_N$, we give
precise definitions and formulate the proposed bound
$s^*_{\mathrm{cyl}}(M,N)\leq9$. The proof of this additional
statement is deferred and is not part of the established results
of the present draft.

An open problem is whether the universal upper bound eight is
sharp for sufficiently large grids with both dimensions at least
ten. Strengthening the
geometric lower bounds and determining $s^*(M,N)$ for wider
grids are natural next steps.

\section*{Acknowledgements}
The authors are delighted to contribute to this special issue honoring Gyula O. H. Katona on the occasion of his 85th birthday and recognizing his fundamental contributions to combinatorial search. Christian Deppe is particularly grateful for Katona's kindness, encouragement, and friendship over many years. He fondly remembers Katona's numerous visits to Bielefeld and deeply appreciates his exceptional commitment to supporting and inspiring young researchers at the Alfréd Rényi Institute of Mathematics. His generosity toward the younger generation has been as inspiring as his outstanding mathematical achievements. V.~Lebedev also has very special memories of Gyula Katona. His first academic visit abroad took him to the Alfréd Rényi Institute of Mathematics to meet Katona. This marked the beginning of a long-standing scientific relationship, with several further visits to Katona in Budapest. Katona, in turn, made several visits to the Institute for Information Transmission Problems of the Russian Academy of Sciences (Kharkevich Institute, IITP), where V.~Lebedev was working at the time. These exchanges remain a particularly cherished part of their scientific collaboration and personal friendship.

The research of A.~Lebedev and V.~Lebedev was supported by the MSHE RF GZ project.
Christian Deppe was supported in part by the German Federal Ministry of Research, Technology and Space (BMFTR) through the projects 6G-life (Grants 16KIS2414 and 16KIS2415) and QSTARS (Grant 16KIS2602).

\section*{Declaration of generative AI and AI-assisted technologies in the manuscript preparation process}
During the preparation of this work, the authors used ChatGPT
(OpenAI) to improve language and grammar
and to assist in writing Python verification programs. The authors
take full responsibility for the final content of the manuscript.

\bibliographystyle{elsarticle-num}
\bibliography{reference}

\end{document}